\input{aipcheck}

\documentclass[
    ,final            
  ]
  {aipproc}

\layoutstyle{6x9}

\begin{document}

\title{Experimental Challenges of the N* Program}

\classification{13.40.Gp, 13.60.Le, 14.20.Gk.}
\keywords      {Baryon structure, constituent quark structure, strong
  degrees of freedom.} 

\author{R.W.~Gothe}{
  address={University of South Carolina, Department of Physics and
  Astronomy, Columbia, SC 29208, USA} 
}

\begin{abstract}
The first challenge faced in investigating the strong interaction from
partially explored, where meson-cloud degrees of freedom dominate, to still
unexplored distance scales, where the dressed-quark contributions are
the dominating degrees of freedom, is to find an experiment that
allows to measure observables that are probing this evolving
nonperturbative QCD regime over the full range. Baryon spectroscopy
can establish more sensitively, and in an almost model-independent
way, nucleon excitation and non-resonant reaction amplitudes by
complete measurements of pseudo-scalar meson photoproduction off
nucleons. Elastic and transition form factors can then trace this
evolution by measurements of elastic electron scattering and exclusive
single-meson and double-pion electroproduction cross sections off the
nucleon that will be extended to higher momentum transfers with the
energy-upgraded CEBAF beam at JLab to study the dressed quark degrees
of freedom, where their strong interaction is responsible for the
ground and excited nucleon state formations. After establishing
unprecedented high-precision data, the immanent next challenge is a
high-quality analysis to extract these relevant electrocoupling
parameters for various resonances that then can be compared to
state of the art models and QCD-based calculations. Recent results
demonstrate the status of the analysis and pinpoint further challenges,
including those to establish QCD-based results directly from the
experimental data.  
\end{abstract}

\maketitle

\section{Introductory Challenges}

\begin{figure}[ht!]
  \includegraphics[height=.45\textheight]{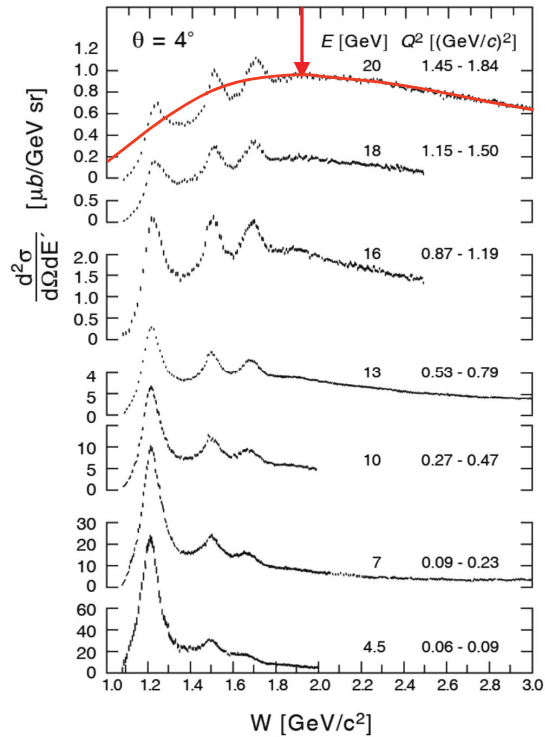}
  \caption{Experimental cross section values versus missing mass for
  early deep inelastic scattering experiments \cite{Ste75}. Sketched
  and marked at $E=20\,GeV$ the partonic quasi-free electron
  scattering peak.}
  \label{dis}
\end{figure}

Already in the early inclusive high-energy deep inelastic scattering (DIS)
experiments at SLAC \cite{Ste75}, scaling and quasi-free scattering off
the constituent quarks became visible at the then highest beam energies of
up to $E=20\,GeV$ but still moderate four-momentum transfers of $Q^2 <
2\,(GeV/c)^2$. In Fig.~\ref{dis} the quasi-free peak becomes visible
at high beam energies and high center-of-mass energies $W$, where
the electrons scatter of constituent quarks. From the marked maximum
value of the quasi-free peak, the constituent quark mass $m_{q} \approx
0.36\,GeV$  and the expectation value of the quark core radius or the 
confinement radius $r_{c} \approx 0.79\,fm$ can be estimated in direct
analogy to the bound-nucleon mass and the nuclear radius in electron
scattering off nuclei, both based on the underlying Fermi
statistics. Figure~\ref{dis} also shows how the quasi-free scattering
starts to dominate the elastic and resonance contributions with
increasing $E$ and $Q^2$. Mapping out this transition over $Q^2$ and
$W$ in detail generates the experimental foundation to investigate
quark-hadron duality \cite{BG70}, scaling, the bound-quark structure,
confinement, dynamical mass generation, and the structure of baryons. The
accepted research proposal PR-09-003 at JLab \cite{GM09}, Nucleon Resonance
Studies with CLAS12, will lay this experimental foundation to address
in a unique way these most challenging questions in QCD. Properly
extracting and interpreting the results from the measured electron
scattering data, particularly for transition form factors to specific
excited nucleon states, might even pose a greater challenge than the 
measurement itself. More measured data on baryon spectroscopy and elastic
form factors are needed to ease this process. A steadily growing
collaboration of experimentalists and theorists is working together to
enable the measurements, the analysis of the data, and the QCD-based
interpretation of the results. The corresponding theoretical progress
and challenges have last been documented in \cite{Az09} and
summarized in this conference \cite{Rob11}. This overview tries to
highlight the experimental challenges of the N* program and does not
even attempt to be complete, but within the productive and well-organized
NSTAR 2011 workshop the status of the N* program was discussed in
detail, which should also be reflected in these proceedings. \\

\section{Spectroscopic Challenges}

\begin{figure}[ht!]
  \includegraphics[height=.4\textheight]{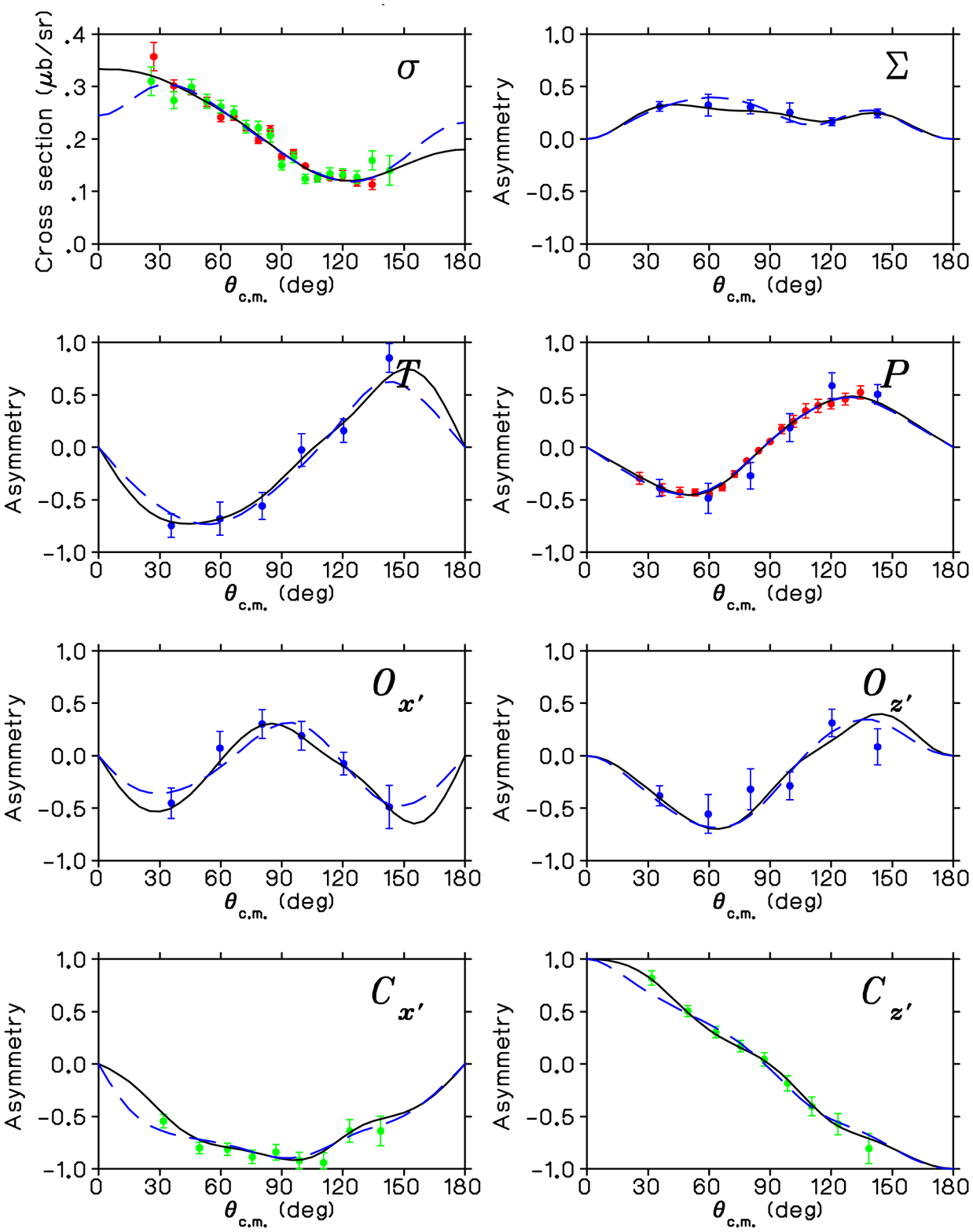}
  \includegraphics[height=.394\textheight]{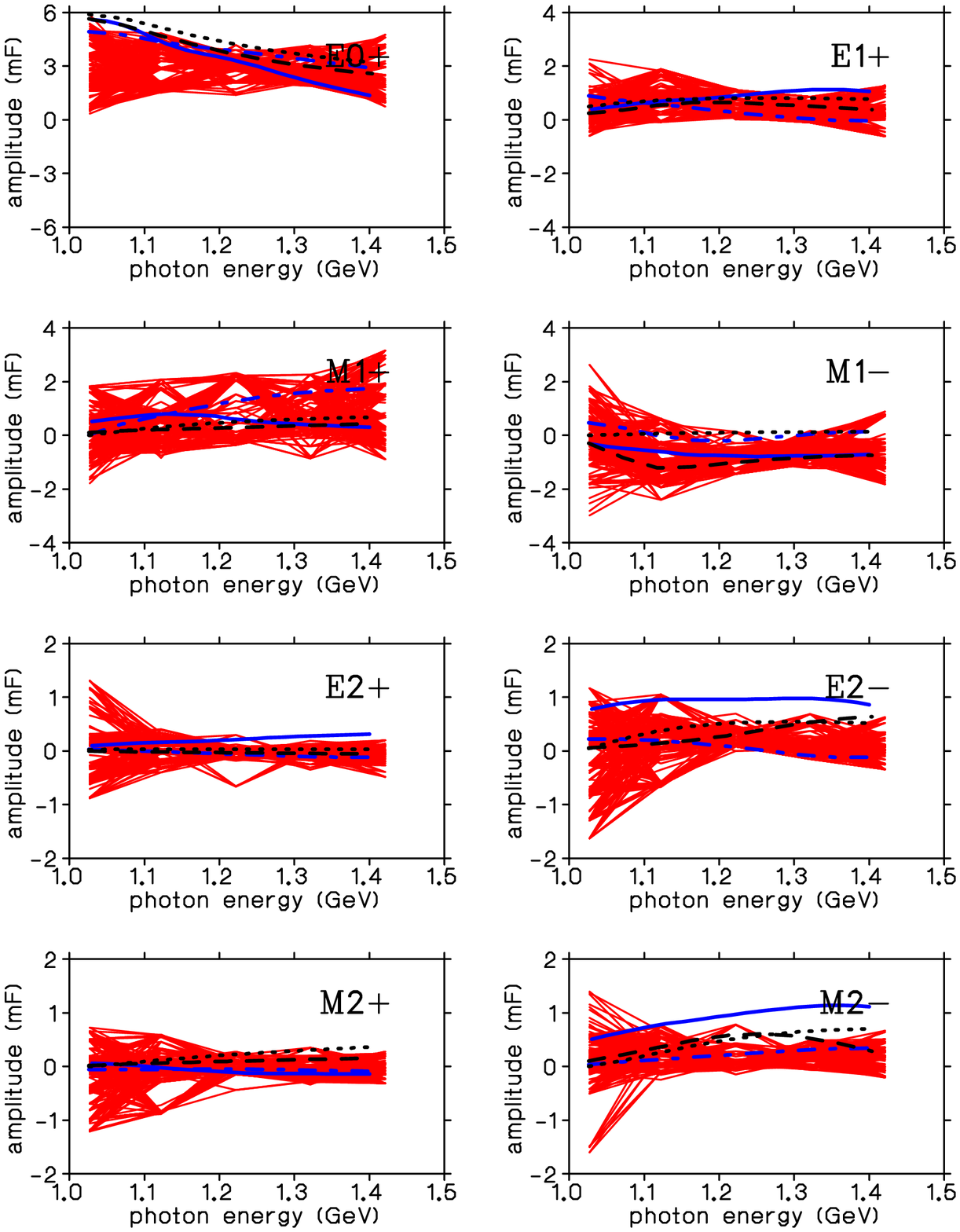}
  \caption{Left side: measured polarization observable data points
  $\sigma$, $P$ CLAS-g11a \cite{MC10} (red), $\Sigma$, $T$, $P$, $O_{x'}$,
  $O_{z'}$ GRAAL \cite{Ll07} (blue), and $\sigma$, $C_{x'}$, $C_{z'}$
  CLAS-g1c \cite{Br07} (green) are compared to the largest (dashed) and best
  $\chi^{2}/d.p.$ fit (solid line) for $W=1883\,MeV$. Right side:
  solution bands (red) for the real parts of lowest lying PWA
  multipoles are compared to the BoGa \cite{BG10} (dash-dotted), Kaon-MAID
  \cite{KM11} (dashed), SAID \cite{SAID} (dotted) and JLST \cite{Ju06} (solid
  lines). \cite{SHKL11}}
\label{shkl}
\end{figure}

The success of quantum electrodynamics was primarily based on the
measurement of the spectrum of the excited hydrogen states, where hydrogen
is the simplest atom bound by electromagnetic fields, whose ground state
can be unambiguously described by its spectrum of excited states. In
analogy one was hoping that by measuring the excitation spectrum of the
simplest baryon bound by strong fields, one could understand the 
structure of the nucleon; and indeed, a large portion of the nuclear
physics community enthusiastically started to investigate baryon
resonances as new optimized detector systems with large solid angle
and momentum coverage and new high-intensity continuous electron beams
became available at many facilities as ELSA, GRAAL, JLab, LEGS, MAMI,
and others world wide. The high versatility
of the provided electromagnetic probe, which has negligible initial
state interaction, has produced intriguing results ever since. It was
realized that the isoscalar or isovector and the electric, magnetic,
or longitudinal nature of the coupling to hadronic matter probes
different aspects of the strong interaction. However, the desired
versatility of the electromagnetic probe comes with the complication
that it mixes all these different coupling amplitudes, each with
various resonance and background contributions, simultaneously into
the measured cross sections. \\
The best possible approach to disentangle all these interfering
amplitudes is to carry out so-called complete experiments. In the
simplest case of pseudo-scalar meson photoproduction, the cross
section can be decomposed into four gauge- and Lorentz-invariant
complex amplitudes. In a combination of unpolarized, beam-, target-,
and recoil-polarization experiments, a total of up to 16 observables can
be measured, of which only eight can be linearly independent. 
Particularly helpful here is that some weak hyperon decays have large
analyzing power, which can replace the explicit recoil polarization
measurement. Large solid angle detectors with high angular resolution
are needed to separate the polarization observables by the azimuthal-
and the partial waves by the polar-angle cross section
distributions. Limited coverage, precision, and statistics will 
restrict the accuracy of the partial wave analysis (PWA) multipoles
more and more severely with growing relative angular momentum $l$,
demanding, even in the simplest and best possible case of complete
experiments, truncation of higher waves or other model-dependent
assumptions. \\
In a recent topical review \cite{SHKL11} the progress in
the analysis of complete experiments is exemplified by the multipole
extraction for the $\gamma p \rightarrow K^{+} \Lambda$
reaction. Figure~\ref{shkl} shows a typical fit to the polar-angle
distribution of the eight measured polarization observables and the
corresponding solution bands for the real parts of the lowest partial
wave multipoles. The width of the solution band depends not only on the
uncertainty of the experimental data point, and hence again on the
angular coverage, precision, and statistics of the experiment, but
also on the number of measured polarization observables. The authors
conclude that based on the eight measured polarization observables the
predicted $D_{13}(1895)$ resonance contributions to the $E_{2-}$ and
$M_{2-}$ PWA multipoles can be excluded, while $P_{13}$ contributions
to $E_{1+}$ and $M_{1+}$ can still not be validated in this $W$
range. This situation will change when data for all 16 polarization
observables within expected uncertainties become available, which
would reduce the solution bandwidth for all shown multipoles (see
Fig.~\ref{shkl}) by roughly a factor of two, allowing to pinpoint the
nature of new resonance contributions in the $K^{+} \Lambda$
channel. \\
With the caveat that most baryon resonances, except the lowest
lying ones, decay dominantly into vector-meson or multi-meson channels, 
complete experiments on single pseudo-scalar meson photoproduction and
PWA will allow for the highest quality extraction
of resonance parameters under minimal model assumptions. 

\section{Hadronic Structure Challenges}

\begin{figure}[t!]
  \includegraphics[height=.30\textheight]{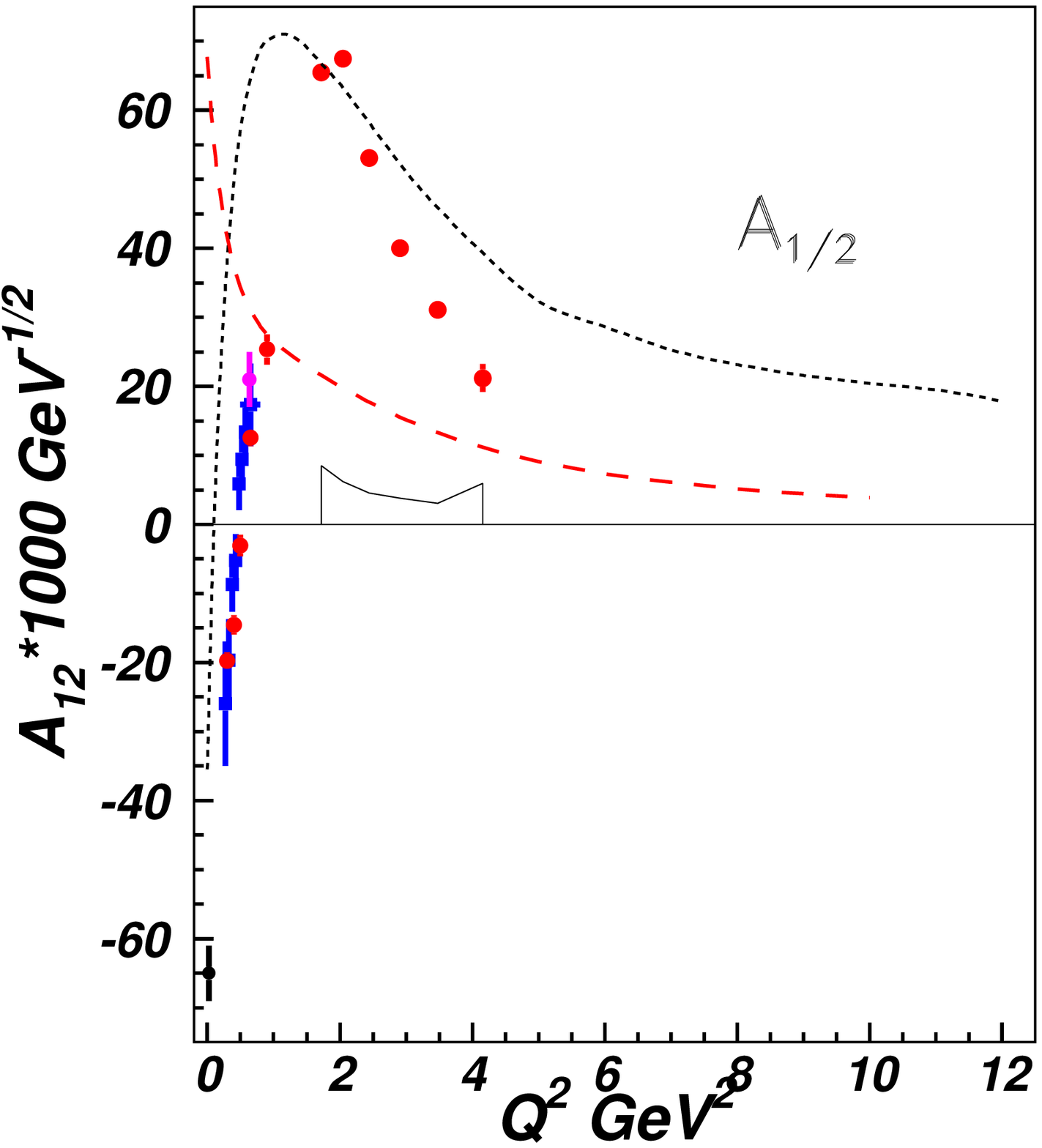}
  \includegraphics[height=.30\textheight]{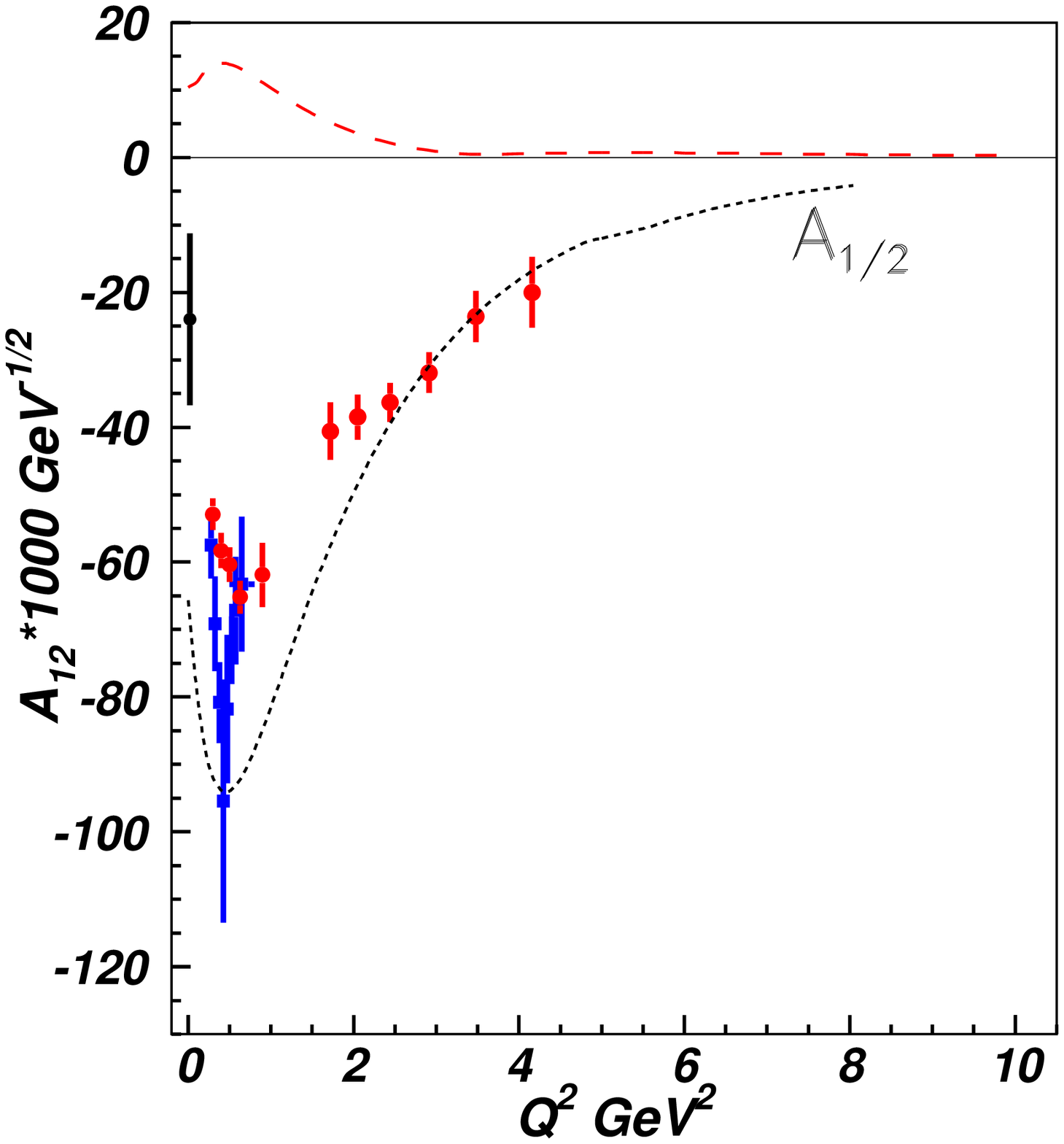}
  \caption{The contributions from quark degrees of freedom and
  meson-baryon dressing to the $A_{1/2}$ electrocouplings of the
  $P_{11}(1440)$ (left) and $D_{13}(1520)$ (right) states. The
  CLAS data from analyses of $N\pi$ \cite{AzB09} and $\pi^{+}\pi^{-}p$
  \cite{Mo09} electroproduction are shown in red and  blue,
  respectively. The contributions from the quark core, estimated within
  the framework of relativistic quark models \cite{Az07,Gia08}, are
  given by the dotted lines, and the absolute value of the
  contributions from meson-baryon 
  dressing, obtained within the framework of the EBAC-DCC approach
  \cite{Lee08}, are represented by the red (dashed) lines.}
\label{ebac}
\end{figure}

\begin{figure}[ht!]
  \includegraphics[height=.27\textheight]{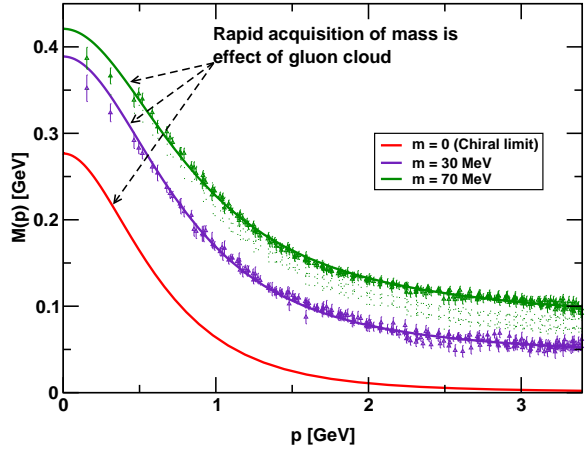}
  \caption{Dressed quark mass function, $M(p)$, for light-quarks,
  obtained in Landau gauge: solid curves are the DSE results,
  including the chiral-limit \cite{Bh03}; points with error bars
  are the results from unquenched LQCD \cite{Bow}. The elastic and
  transition form factor data, that will become available after the
  $12\,GeV$ upgrade, probes $M(p)$ up to a quark propagator
  momentum of $1.15\,GeV/c$, which spans the transition from
  dressed constituent quarks to the
  almost-completely undressed current quarks. }
\label{rqmass}
\end{figure}

Beyond baryon spectroscopy at the real photon point
$Q^2=0\,(GeV/c)^2$, electron scattering experiments also investigate
the internal hadronic structure at various distance scales by tuning
the four-momentum transfer from $Q^2 \approx 0\,(GeV/c)^2$, where the
meson cloud contributes significantly to the baryon structure, over
intermediate $Q^2$, where the three constituent-quark core starts to
dominate, to $Q^2$ up to $12\,(GeV/c)^2$, attainable after the
$12\,GeV$ upgrade at JLab, where the constituent quark
gets more and more undressed towards the bare current quark (see
Fig.~\ref{rqmass}) \cite{Az11,Bh03}. 
Although originally derived in the high $Q^2$
limit, constituent counting rules describe in more general terms how
the transition form factors and the corresponding helicity amplitudes
scale with $Q^2$ dependent on the number of effective
constituents. Recent results for $Q^2 < 5\,(GeV/c)^2$
\cite{AzB09,Par08} indicate for some helicity amplitudes, like
$A_{1/2}$ for the electroexcitation of the $N(1520)D_{13}$,
the onset of proper scaling assuming three constituent quarks. This
further indicates that in this case the meson-baryon contributions
become negligible in comparison to those of a three constituent-quark
core, which coincides nicely with the EBAC dynamical coupled channel
calculation shown in Fig.~\ref{ebac} (right). \\
Along the same line of reasoning perturbative QCD (pQCD) predicts in
the high $Q^2$-limit by neglecting higher twist contributions that
helicity is conserved. The fact that this predicted behavior sets in
at much lower $Q^2$ values than expected \cite{AzB09} challenges our current
understanding of baryons even further. For $N(1520)D_{13}$ the
helicity conserving amplitude $A_{1/2}$ starts to dominate the
helicity non-conserving amplitude $A_{3/2}$ at $Q^2 \approx
0.7\,(GeV/c)^2$, as typically documented by the zero crossing of the
corresponding helicity asymmetry    
$ A_{hel} = (A_{1/2}^2 - A_{3/2}^2)/(A_{1/2}^2 + A_{3/2}^2)$.
The $N(1685)F_{15}$ resonance shows a similar behavior with a zero
crossing at $Q^2 \approx 1.1\,(GeV/c)^2$, whereas the
$\Delta(1232)P_{33}$ helicity asymmetry stays negative with no
indication of an upcoming zero crossing; and even more surprising are
the preliminary results for the $N(1720)P_{13}$ $A_{1/2}$ amplitude,
which decreases so rapidly with $Q^2$ that the helicity asymmetry
shows an inverted behavior with a zero crossing from positive to
negative around $Q^2 \approx 0.7\,(GeV/c)^2$ \cite{Mok10}.  
This essentially different behavior of resonances underlines that it
is necessary but not sufficient to extend the measurements of only the
elastic form factors or the transition form factors to a well known
resonance like the $\Delta(1232)P_{33}$ to higher momentum transfers. 
To comprehend QCD at intermediate distance scales where dressed quarks
are the dominating degrees of freedom and to explore interactions of
dressed quarks as they form various baryons in distinctively different
quantum states, the $Q^2$ evolution of exclusive transition form
factors to various resonances up to $12\,(GeV/c)^2$ are absolutely
crucial. \linebreak
Figure~\ref{ebac} exemplifies the three corner stones of the status quo
in this baryonic structure analysis endeavor. First, the analysis of the $N\pi$
channel data is carried out in two phenomenologically different approaches
based on fixed-$t$ dispersion relations and a unitary isobar
model~\cite{AzB09,Az03}. The main difference between the two
approaches is the way the non-resonant contributions are derived. The
$p\pi^+\pi^-$  CLAS data is analyzed within a phenomenological meson-baryon
model \cite{Mo09,MoB09} that fits nine independent differential
cross sections of invariant masses and angular distributions. The good
agreement of the resonant helicity amplitude results in the single-
and double-pion channels, that have fundamentally different
non-resonant contributions, provides evidence for the reliable
extraction of the $\gamma_vNN^*$ electrocoupling amplitudes. 
Second, the high $Q^2$ behavior is most consistently described by
relativistic light-front quark models, like
\cite{Az07,Gia08,Cap95,Web90}, but their description of the 
low $Q^2$ behavior is less satisfactory. Third, the Excited 
Baryon Analysis Center (EBAC) predicts, based on a full dynamical
coupled-channel analysis \cite{Lee07}, meson-baryon dressing (meson-cloud)
contributions that seem to bridge the gap \cite{Az09} between the
relativistic light-front quark models and the measured results at low
$Q^2$. \\ 
Digging deeper into the baryonic structure by increasing the momentum
transfer beyond $5\,(GeV/c)^2$ \cite{GM09} opens a unique window to
investigate the dynamic momentum-dependent structure of the
constituent quarks. This becomes apparent in Fig.~\ref{rqmass}, where
the quark mass function 
for momenta larger than $2\,GeV/c$ describes a current-quark that
propagates almost like a free single parton. However, for momenta less
than that, the quark mass function rises sharply, entering the
confinement regime and reaching the constituent-quark mass scale in
the infrared.  In this domain, the quark is far from being a
single parton and is dressed by a cloud of low-momentum gluons
attaching themselves to the current-quark, which is a direct
manifestation of dynamical chiral-symmetry breaking. 

\section{Summary}

All visible matter that surrounds us is made of atoms, which are made
of electrons and nuclei; the latter are made of nucleons, which are
finally made of quarks and gluons. Contrary to the most recent
discussions in the news, the Higgs, or more frequently called the God
Particle, is not responsible for the generation of all mass. 
It is already known that 98\% of
all visible mass is generated by strong fields. Establishing an
experimental and theoretical program that provides access to
 \begin{itemize}
\item  the dynamics of nonperturbative strong interactions among
dressed quark, their emergence from QCD, and their confinement in
baryons, 
\item the dependence of the light quark mass on the momentum transfer and
thereby how the constituent quark mass arises from dynamical
chiral-symmetry breaking, and 
\item the  behavior of the universal QCD $\beta$-function in the
infrared regime, 
\end{itemize}
is indeed most challenging on all levels, but recent progress and future
commitments bring a solution of these most fundamental remaining QCD
problems into reach. \\
Single-, double-, and triple-polarization
experiments are essential to establish anchor points for the most
detailed separation of individual resonance and background
contributions. Elastic and particularly transition form factors are
then uniquely accessing nonperturbative QCD from long to short
distance scales. 

\section{Acknowledgments}
This work was supported in part by the National Science Foundation,
the U.S. Department of Energy, and other international funding agencies
supporting research groups at Jefferson Lab.

\bibliographystyle{aipproc}   

\bibliography{sample}

\IfFileExists{\jobname.bbl}{}
 {\typeout{}
  \typeout{******************************************}
  \typeout{** Please run "bibtex \jobname" to optain}
  \typeout{** the bibliography and then re-run LaTeX}
  \typeout{** twice to fix the references!}
  \typeout{******************************************}
  \typeout{}
 }

\end{document}